\documentclass{myelsart} 
\usepackage{graphics}
\usepackage{amssymb} 
\usepackage{amsmath}

\begin{document}

\begin{frontmatter}

\title{Transition States in Protein Folding Kinetics: The Structural Interpretation of $\Phi$-values}
\author{Thomas R.\ Weikl$^1$ and Ken A.\ Dill$^2$}
\address{$^1$Max Planck Institute of Colloids and Interfaces, Theory Department,\\14424 Potsdam, Germany\\[0.1cm]
$^2$Department of Pharmaceutical Chemistry, University of California, \\ San Francisco, California 94143-2240, USA} 

\begin{abstract}
$\Phi$-values are experimental measures of the effects of mutations on the folding kine\-tics of a protein. A central question is which structural information $\Phi$-values contain about the transition state of folding. Traditionally, a $\Phi$-value is interpreted as the `nativeness' of a mutated residue in the transition state.  However, this interpretation is often problematic because it assumes a linear relation between the nativeness of the residue and its free-energy contribution. We present here a better structural interpretation of $\Phi$-values for mutations within a given helix. Our interpretation is based on a simple physical model that distinguishes between secondary and tertiary free-energy contributions of helical residues. From a linear fit of our model to the experimental data, we obtain two structural parameters: the extent of helix formation in the transition state, and the nativeness of tertiary interactions in the transition state.  We apply our model to all proteins with well-characterized helices for which more than 10 $\Phi$-values are available: protein A, CI2, and protein L. The model captures nonclassical $\Phi$-values $<0$ or $>1$ in these helices, and explains how different mutations at a given site can lead to different $\Phi$-values.  
\end{abstract}
\end{frontmatter}

\section*{Introduction}

There has been much interest in understanding the rates of protein folding in terms of transition state structures.  We focus here on two-state proteins, i.e.\ those proteins that fold with single-exponential kinetics. The folding kinetics of two-state proteins is often investigated by mutational analysis \cite{Itzhaki95,Villegas98,Chiti99,Ternstrom99,Fulton99,Kim00,Mccallister00,Otzen02,Hedberg04,Anil05,Went05,Wilson05,Kragelund99,Gianni03,Sato04,Teilum05,Martinez99,Riddle99,Hamill00,Fowler01,Cota01,Jaeger01,Northey02,Garcia04}. The effect of a given mutation on the protein's folding kinetics is quantified by its $\Phi$-value \cite{Matouschek89,Fersht99}
\begin{equation}
\Phi= \frac{R T\ln( k_{\text{wt}}/k_{\text{mut}})}{\Delta G_{N}} \label{phi}
\end{equation}
Here, $k_{\text{wt}}$ is the folding rate for the wildtype protein, $k_{\text{mut}}$ is the folding rate for the mutant protein, and $\Delta G_{N}$ is the change of the protein stability induced by the mutation. The stability $G_{N}$ of a protein is the free energy difference between the denatured state $D$ and the native state $N$. In classical transition-state theory, 
the folding rate of a two-state protein is proportional to $\exp[-G_T /RT]$, where $G_T$ is the free energy difference from the denatured state to the transition state.\footnote{In principle, the prefactor of this proportionality relation could also depend on the mutation, but this dependence is usually neglected.} In that notation, $\Phi$-values have the form 
\begin{equation}
 \Phi=\frac{\Delta G_{T}}{\Delta G_{N}}  \label{phi2}
\end{equation} 
where each $\Delta$ in this expression represents the change due to the mutation.  

By definition, $\Phi$-values are energetic quantities, related to changes in the protein's stability and folding rate. Do $\Phi$-values also give information about the structures that the protein adopts when it is in a kinetic ``bottleneck'' or transition state \cite{Fersht99,Ozkan01,Zarrine04,Chang04,Raleigh05}? In the traditional interpretation, $\Phi$-values are taken to indicate the {\em degree of structure formation of the mutated residue in the transition-state ensemble T.} A  $\Phi$-value of 1 is interpreted to indicate that the residue is fully native-like structured in T, since the mutation shifts the free energy of the transition state T by the same amount as the free energy of the native state N. A $\Phi$-value of 0 is interpreted to indicate that the residue is as unstructured in T as in the denatured state D, since the mutation does not shift the free energy difference between these two states. $\Phi$-values between 0 and 1 are taken to indicate partial native-like structure in T.   

Modelers often calculate $\Phi$-values based on this traditional interpretation. In many approaches, $\Phi$-values are calculated from the fraction of contacts a residue forms in the transition state T, compared to the fraction of contacts in the native and the denatured  state \cite{Li94,Li96,Lazaridis97,Vendruscolo01,Li01,Gsponer02,Paci02,Guo03,Settanni04,Paci05,Salvatella05,Chong05,Hubner05,Duan05}, or from similar structural parameters \cite{Daggett96,Day05}. Notable exceptions are a recent MD study of an ultrafast mini-protein in which $\Phi$-values are calculated from rates for the wildtype and mutants via eq.~(\ref{phi}) \cite{Settanni05}, and the calculation of $\Phi$-values from free energy shifts of the transition-state ensemble using eq.~(\ref{phi2}) \cite{Lindorff03}. 

However, there are reasons to question this simple interpretation of $\Phi$-values.  First, some $\Phi$-values are negative or larger than 1 \cite{Goldenberg99,Rios05}. These `nonclassical' $\Phi$-values cannot be interpreted as a degree of structure formation, because this would have the nonsensical implication of `less structured than $D$' or `more structured than $N$'. Second, $\Phi$-values are sometimes significantly different for different mutations at a given chain position, contradicting the normal assumption that the degree of nativeness of the transition state is just a property of the position of a monomer in the protein.  Third, $\Phi$-values for neighboring residues within a given secondary structure often span a wide range of $\Phi$-values. In the traditional interpretation, this means that some of the helical residues are unstructured in the transition state, while other residues, often direct neighbors, are highly structured. This contradicts the notion that secondary structures are cooperative. 

The inconsistencies of the traditional interpretation result from the assumption that the mutation-induced free energy changes of a residue are proportional to a single structural parameter, the `degree of nativeness' of this residue in the transition state T. Is there a consistent structural interpretation of $\Phi$-values, and if yes, how many structural parameters do we need to capture the mutation-induced free energy changes? We show here that the $\Phi$-values for multiple mutations in a given helix can be consistently interpreted in a simple physical model that takes into account just two structural parameters for the whole helix: $\chi_\alpha$, the  {\em degree of secondary structure formation of the helix in the transition-state ensemble T}, and $\chi_t$ the {\em degree of tertiary structure formation of the helix in T}. In our model, the mutation-induced free energy changes are split into two components. The overall stability change $\Delta G_N$ is split into two parts: the change in intrinsic helix stability $\Delta G_\alpha$, and the change in tertiary free energy $\Delta G_t$ caused by the mutation. Similarly, $\Delta G_T$, the change of the free energy difference between the transition state and the denatured state, is split into a change $\chi_\alpha\Delta G_\alpha$ in secondary free energy, and a change  $\chi_t\Delta G_t$ in tertiary free energy. The $\Phi$-values for the mutations in the helix then have the general form
\begin{equation}
\Phi = \frac{\chi_\alpha\Delta G_\alpha + \chi_t\Delta G_t}{\Delta G_N} = \chi_t + \left(\chi_\alpha - \chi_t\right)\frac{\Delta G_\alpha}{\Delta G_N} \label{phi_modeled}
\end{equation}
The second expression simply results from replacing $\Delta G_t$ by $\Delta G_N - \Delta G_\alpha$. The two parameters $\chi_\alpha$ and $\chi_t$ of our model are `collective' structural parameters for all mutations in the helix. Different $\Phi$-values then simply result from different free-energetic `signatures' $\Delta G_\alpha$ and $\Delta G_N$ of the mutations. In particular, eq.~(\ref{phi_modeled}) captures that different mutations of the same residue can lead to different $\Phi$-values, and that $\Phi$-values can be `nonclassical', i.e.~$<0$ or $>1$.  Since the two structural parameters $\chi_\alpha$ and $\chi_t$ range between 0 and 1, a nonclassical $\Phi$-value implies that the changes $\Delta G_\alpha$ and $\Delta G_t$ in secondary and tertiary free energy caused by the mutation have opposite signs.  

To apply our model, we first estimate $\Delta G_\alpha$, the change in helical stability, for each mutation in a particular helix, using standard helix propensity methods. We then plot all experimental values for $\Phi$ versus  $\Delta G_{\alpha}/\Delta G_N$, and obtain the two structural parameters $\chi_\alpha$ and $\chi_t$ from a linear fit of eq.~(\ref{phi_modeled}). In principle, the two structural parameters can be extracted if $\Phi$-values and stability changes for at least two mutations in a helix are available. However, to test our model, and to obtain reliable values for $\chi_\alpha$ and $\chi_t$, we focus here on helices for which more than 10 $\Phi$-values have been determined. The modeling quality then can be assessed from the standard deviation of the data points from the  regression line, and from the Pearson correlation coefficients between  $\Phi$ and  $\Delta G_{\alpha}/\Delta G_N$. Our model can be applied to all mutations for a helix, or to a subset of mutations that affect only the tertiary interactions with one other structural element. 

\section*{Models and methods}

\subsection*{Transition-state conformations and folding rate} 

\begin{figure}
\begin{center}
\resizebox{0.5\linewidth}{!}{\includegraphics{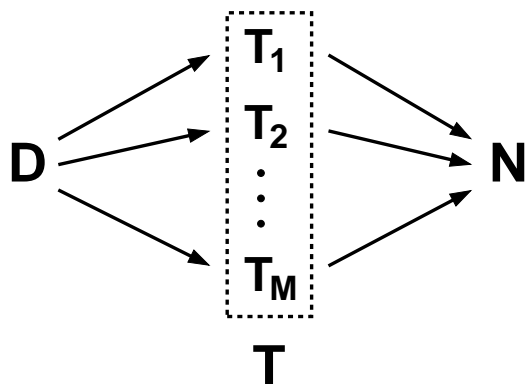}}
\end{center}
\vspace{1cm}
\caption{In our model, the transition-state ensemble T consists of $M$ transition-state conformations T$_1$, T$_2$, $\ldots$, T$_M$. The arrows indicate the folding direction from the denatured state D to the native state N via the transition-state conformations.}
\label{figure_ts}
\end{figure}

We model the transition state as an ensemble of $M$ different conformations (see Fig.~\ref{figure_ts}). Each transition-state conformation is directly connected to the native state N and to the denatured state D. The model thus has $M$ parallel folding and unfolding routes.

We assume that the protein is stable, i.e.\ that $G_N < 0$.  We also assume that the free energy barrier for each transition state conformation is significantly larger than the thermal energy, i.e.\ that $G_m/RT  \gg 1$ \cite{Schuler02,Akmal04}. The rate of folding along each route $m$ is then proportional to $\exp[-G_m/RT]$, and the total folding rate as the sum over all the parallel routes is  
\begin{equation}
k = c \sum_{m=1}^M e^{-G_m/RT}    \label{eq_foldingrate}
\end{equation}
where $c$ is a constant prefactor.\footnote{This model is a generalization of our previous model \cite{Merlo05} with $M=2$ transition-state conformations. The master equation that describes the folding kinetics of this model can be solved exactly. Eq.~(\ref{eq_foldingrate}) is obtained from the exact solution in the limit of large transition state barriers $G_m$\cite{Merlo05}.}

\subsection*{Decomposition of free energy changes for helical mutations}
 
Consider all mutations $i=1,2,\ldots$ within one particular $\alpha$-helix of a protein. The effect of these mutations on the stability and folding kinetics can be experimentally characterized by the stability changes $\Delta G_N$, and by the $\Phi$-values. We suppose that the experimentally measured change in stability $\Delta G_N$ for each mutation is the sum of effects on the stability of the helix and on the interactions of the helix with its tertiary neighbors:
\begin{equation}
\Delta G_N=\Delta G_\alpha + \Delta G_t  
\end{equation}
The first term, $\Delta G_\alpha$, is the change in the intrinsic helix stability. The \mbox{second} term, $\Delta G_t$, is the change in tertiary free energy of the helix interactions with neighboring structures. Below, we estimate $\Delta G_\alpha$ using either the  program AGADIR \cite{Munoz94a,Munoz94b,Lacroix98} or from a helix propensity scale \cite{Pace98}. The term $\Delta G_t$ is then simply obtained by subtracting $\Delta G_\alpha$ from the experimentally measured stability change, $\Delta G_N$.

We also decompose each $\Delta G_m$, the mutation-induced free energy change for the transition state conformation $m$, into two terms:
\begin{equation}
\Delta G_m = s_m \Delta G_\alpha + t_m \Delta G_t \label{eq_mutantcomponents}
\end{equation}
Here, $s_m$ is either 0 or 1, depending on whether the helix is formed or not in the transition state conformation $m$. The coefficient $t_m$ is between 0 and 1 and represents the degree of tertiary structure formation in conformation $m$. 
 
\subsection*{Structural and energetic components of $\Phi$-values}

The folding rate for the mutant protein $i$ is $k_\text{mut} =   k\big(G_1+\Delta G_1, G_2+\Delta G_2,\ldots, G_M+\Delta G_M\big)$ with $k$  given in eq.~(\ref{eq_foldingrate}). The folding rate of the wildtype is $k_\text{wt}= k(G_1,G_2,\ldots,G_M)$. We assume here that the mutations do not affect the pre\-factor $c$ in eq.~(\ref{eq_foldingrate}). For small values $\left|\Delta G_m\right|$ of the mutation-induced free-energy changes, a Taylor expansion of $\ln k_\text{mut}$ gives
\begin{equation}
\ln k_\text{mut} -\ln k_\text{wt} \simeq  \sum_{m=1}^M \frac{\partial \ln k_\text{wt}}{\partial G_m}\Delta G_m 
= -\frac{1}{RT}\frac{\sum_m  \Delta G_m e^{-G_m/RT}}{\sum_m e^{-G_m/RT}}
\end{equation} 
With the decomposition of the $\Delta G_m$'s in eq.~(\ref{eq_mutantcomponents}), we obtain
\begin{equation} 
\ln k_\text{mut} -\ln k_\text{wt} \simeq  -\frac{1}{RT}\left(\chi_\alpha \Delta G_\alpha + \chi_t \Delta G_t \right) \label{deltalogk}
\end{equation} 
with the two terms
\begin{equation}
\chi_\alpha  \equiv \frac{\sum_m s_m e^{-G_m/RT}}{\sum_m e^{-G_m/RT}} \text{~~and~~}
\chi_t  \equiv \frac{\sum_m s_t e^{-G_m/RT}}{\sum_m e^{-G_m/RT}} ~.
\end{equation}
The term $\chi_\alpha$ represents the Boltzmann-weighted average of the secondary structure parameter $s_m$ for the transition-state ensemble T. $\chi_\alpha$ ranges from 0 to 1 and indicates the average degree of structure formation for the helix in T. The value $\chi_\alpha=1$ indicates that the helix is formed in all transition-state conformations $m$, and $\chi_\alpha=0$ indicates that the helix is formed in none of the transition-state conformations. Values of $\chi_\alpha$ between 0 and 1 indicate that the helix is formed in some of the transition-state conformation, and not formed in others. The term $\chi_t$ represents the Boltzmann-weighted average of the tertiary structure parameter $t_m$ in T, and also ranges from 0 to 1. From eq.~(\ref{deltalogk}) and the definition in eq.~(\ref{phi}), we then obtain the general form (\ref{phi_modeled}) of the $\Phi$-values for helical mutations in our model. 
\footnote{In principle, our parameter $\chi_t$ for the tertiary interactions can also be seen to depend on the residue position. To derive eq.~(\ref{phi_modeled}), we don't have to assume that the tertiary parameters $t_m$ for the $m$ transition-state conformations are independent of the residue position and/or mutation. However, we focus here on the simplest version of our model and show that a consistent structural interpretation of experimental $\Phi$-values in a helix can be obtained with just two structural parameters $\chi_\alpha$ and $\chi_t$ for the whole helix, which implies a cooperativity of secondary as well as tertiary interactions.}

More than twenty two-state proteins with  $\alpha/\beta$ \cite{Itzhaki95,Villegas98,Chiti99,Ternstrom99,Fulton99,Kim00,Mccallister00,Otzen02,Hedberg04,Anil05,Went05,Wilson05}, $\alpha$-helical \cite{Kragelund99,Gianni03,Sato04,Teilum05}, or all-$\beta$ structures \cite{Martinez99,Riddle99,Hamill00,Fowler01,Cota01,Jaeger01,Northey02,Garcia04} have been investigated by mutational analysis in the past few years. Mutational data are also available for several proteins that fold via intermediates \cite{Serrano92,Jemth05,Zhou05} or apparent intermediates \cite{Scott04}. We focus here on the well-characterized $\alpha$-helices of two-state proteins for which at least 10 $\Phi$-values apiece are available: the helices 2 and 3 from the protein A, and the helices of CI2 and protein L. Protein A is an $\alpha$-helical protein with three helices, CI2 and protein L are $\alpha/\beta$-proteins with a single $\alpha$-helix packed against a $\beta$-sheet.

\section*{Results and discussion}

Our analysis of experimental $\Phi$-values requires an estimate of the mutation-induced changes $\Delta G_\alpha$ of the intrinsic helix stability. In the case of the CI2 helix, we estimate $\Delta G_\alpha$ both with the program AGADIR \cite{Munoz94a,Munoz94b,Lacroix98} and from a helix propensity scale \cite{Pace98}, see Table 1. The change in intrinsic helix stability $\Delta G_\alpha$ can be estimated from the helical content predicted by AGADIR via $\Delta G_\alpha= RT \ln \left(P_\alpha^\text{wt}/P_\alpha^\text{mut}\right)$. Here, $P_\alpha^\text{wt}$ is the helical content of the wildtype helix, and $P_\alpha^\text{mut}$ the helical content of the mutant. The program AGADIR is based on helix/coil transition theory, with parameters fitted to data from Circular Dichroism (CD) spectroscopy. In Table 1, the values for  $\Delta G_\alpha$ obtained from AGADIR are compared to values from a helix propensity scale \cite{Pace98}. Helix propensities of the amino acids are typically given as free energies differences with respect to Alanine. We use the propensity scale of Pace and Scholtz \cite{Pace98}, which has been obtained from experimental data on 11 different helical systems. For example, the value  $\Delta G_\alpha=0.29$ kcal/mol for the mutant E15D  in the CI2 helix is simply the difference between the helix propensity 0.69 kcal/mol for the amino acid D (Aspartic acid) and the propensity 0.40 kcal/mol for amino acid E (Glutamic acid). The helix propensity scale can be applied for residues at `inner' positions' of a helix, not for residues at the termini or `caps' of the helix. The N-terminal residues of the CI2 helix are the residues 12 and 13, the C-terminal residues are the residues 23 and 24. For the 8 mutations at `inner positions' of the CI2 helix, the values for $\Delta G_\alpha$ from AGADIR and from the helix propensity scale correlate with a Pearson correlation coefficient of 0.77. For the other three helices considered here, the helicities predicted by AGADIR are significantly smaller than the helicities around 5 \% predicted for the CI2 helix. Estimates for $\Delta G_\alpha$ based on AGADIR therefore are not reliable for these helices. The values of $\Delta G_\alpha$ shown in the Tables 2 to 4 are calculated from helix propensities. 

\begin{figure}
\begin{center}
\resizebox{0.6\linewidth}{!}{\includegraphics{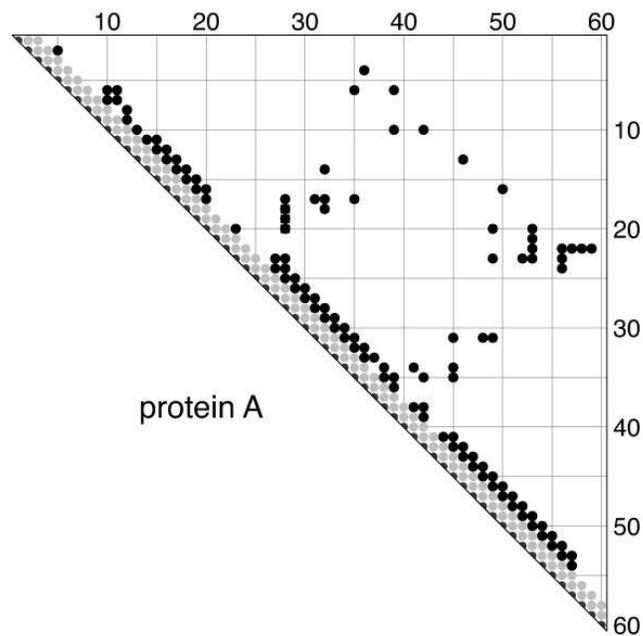}}
\end{center}
\caption{Contact matrix of protein A. A black dot at position $(i,j)$ of the matrix indicates that the two non-neighboring residues $i$ and $j$ are in contact in the native structure (protein data bank file 1SS1, model 1). Two residues here are defined to be in contact is the distance between any of their non-hydrogen atoms is smaller than the cutoff distance 4 \AA. Protein A is an $\alpha$-helical protein with three helices. Helix 1 consists of the residues 10 to 19, helix 2 of the residues 25 to 37, and helix 3 of the residues 42 to 56. 
}
\label{map_proteinA}
\end{figure}

\begin{figure}
\begin{center}
\resizebox{0.6\linewidth}{!}{\includegraphics{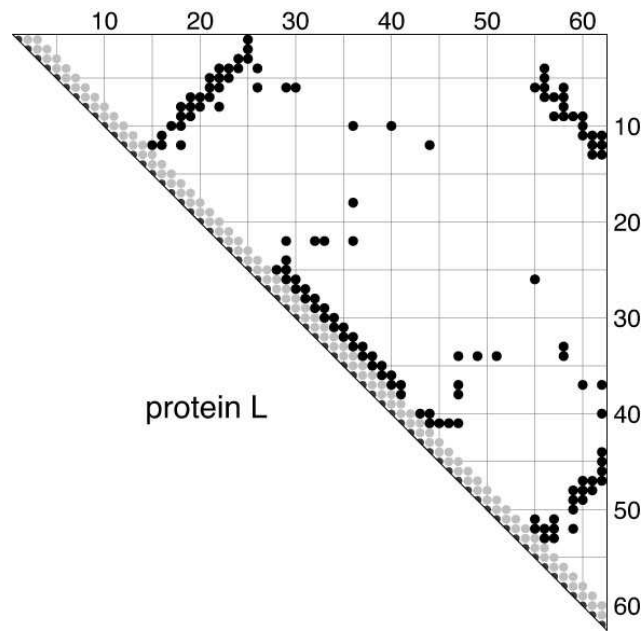}}
\end{center}
\caption{Contact matrix of protein L (protein data bank file 1HZ6, residues A1 to A62). The structure of protein L consists of two $\beta$-hairpins at the termini, and an $\alpha$-helix in between. The helix consists of the residues 26 to 40. The hairpin 1 at the N-terminus includes the residues 4 to 24, and the hairpin 2 at the C-terminus includes the residues 47 to 62.}
\label{map_proteinL}
\end{figure}

The three structural elements of protein A are its three helices. Based on the contact map of protein A shown in Fig.~\ref{map_proteinA}, the mutations in helix 2 of protein A can be divided into three groups: `purely secondary' mutations that don't affect tertiary contacts; mutations that affect only tertiary contacts with helix 1; and mutations that affect tertiary contacts both with helix 1 and 3. If only the first two groups of mutations are considered in our analysis, $\chi_t$ represents the average degree of structure formation with helix 1. If all groups and, thus, all mutations are considered, $\chi_t$ is the average degree of structure formation with the helices 1 and 3.  In the case of helix 3, we distinguish between mutations that affect either tertiary contacts with helix 1 or helix 2, or none of the tertiary interactions, see Table 3. In the case of the protein L helix, the two other structural elements are the terminal $\beta$-hairpins, see Fig.~\ref{map_proteinL} and Table 4. In the case of CI2, we do not distinguish between different tertiary contacts. One reason is that there are at least three other structural elements to consider, the three strand pairings $\beta_2\beta_3$, $\beta_3\beta_4$, and   $\beta_1\beta_4$ of the four-stranded $\beta$-sheet that is packed against the CI2 helix \cite{Merlo05}. Another reason is that the degree $\chi_t$ of tertiary structure formation in the transition state is small for this helix.

\begin{figure}
\begin{center}
\resizebox{0.7\linewidth}{!}{\includegraphics{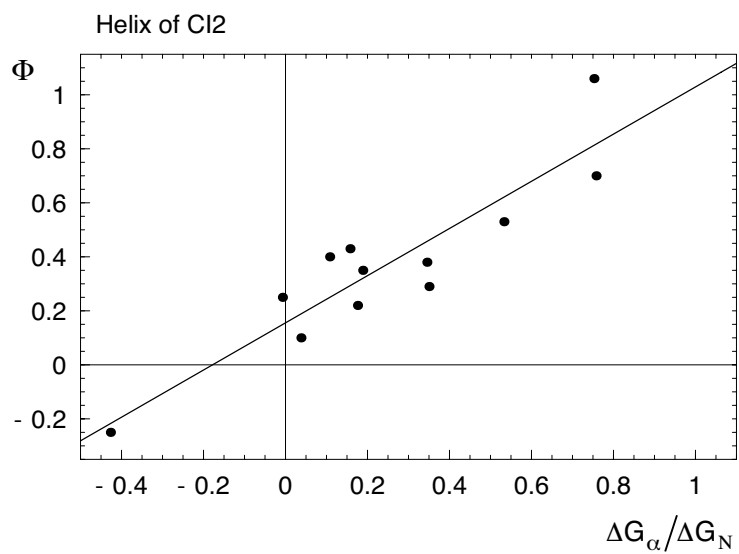}}
\end{center}
\caption{Analysis of $\Phi$-values for mutations in the helix of the protein CI2. The change in intrinsic helix stability $\Delta G_\alpha$ for the 12 mutations has been calculated with AGADIR (see Table 1). We only consider mutations with experimentally measured stability changes $\Delta G_N > 0.7$ kcal/mol. The Pearson correlation coefficient of the 12 data points is 0.91. From the regression line $\Phi = 0.16 + 0.87 \Delta G_\alpha/\Delta G_N$, we obtain the structural parameters $\chi_\alpha = 1.03\pm 0.05$ and $\chi_t= 0.16\pm 0.05$. The structural parameter $\chi_\alpha$ close to 1 indicates that the helix is fully formed in the transition state. The parameter $\chi_t$ indicates that tertiary interactions are on average present in the transition state to a degree around 16 \% }
\label{helix_ci2}
\end{figure}

\begin{figure}
\begin{center}
\resizebox{0.7\linewidth}{!}{\includegraphics{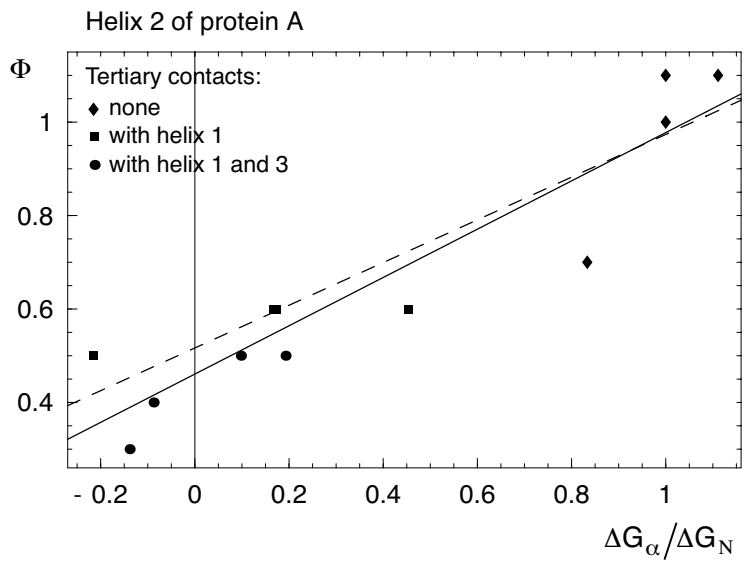}}
\end{center}
\caption{Analysis of $\Phi$-values for helix 2 of protein A. The solid line represents the regression line $\Phi = 0.46 + 0.52 \,\Delta G_\alpha/\Delta G_N$ for all points. The correlation coefficient of the data points is 0.93. The dashed line is the regression line $\Phi = 0.52 + 0.46 \,\Delta G_\alpha/\Delta G_N$ of the 8 data points for mutations of residues that have either no tertiary interactions or tertiary interactions with helix 1 (see also Table 2). The correlation coefficient of these data points is 0.90. From the regression lines and eq.~(\ref{phi_modeled}), we obtain the structural parameters $\chi_\alpha$ and $\chi_t$ shown in Table 5. The values of $\chi_\alpha$ close to 1 indicate that the helix is fully formed in the transition state, and the values of $\chi_t$ close to 0.5 indicate that tertiary interactions are present to a degree of about 50 \%.
}
\label{helix2_proteinA}
\end{figure}

\begin{figure}
\begin{center}
\resizebox{0.7\linewidth}{!}{\includegraphics{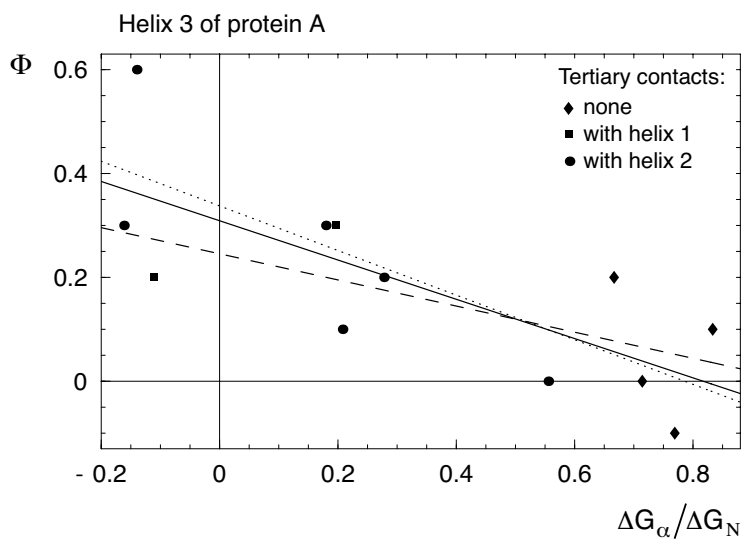}}
\end{center}
\caption{Analysis of $\Phi$-values for mutations in helix 3 of protein A. The solid line represents the regression line $\Phi = 0.31 - 0.38 \Delta G_\alpha/\Delta G_N$ of all data points; the dashed line is the regression line $\Phi = 0.24 - 0.25 \Delta G_\alpha/\Delta G_N$ of the data points for mutations that affect the tertiary interactions with helix 1 (or no tertiary interactions); and the dotted line is the regression line $\Phi = 0.34 - 0.43 \Delta G_\alpha/\Delta G_N$ of data points for mutations that affect tertiary interactions interactions with helix 2 or no tertiary interactions). The absolute values of the correlation coefficient for these three data sets are $|r|=0.75$, 0.65, and 0.79, respectively (see Table 5). 
}
\label{helix3_proteinA}
\end{figure}

\begin{figure}
\begin{center}
\resizebox{0.7\linewidth}{!}{\includegraphics{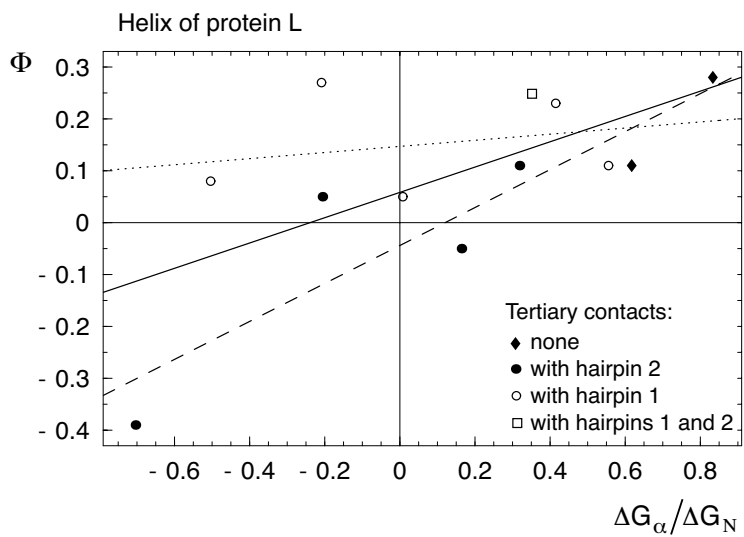}}
\end{center}
\caption{Analysis of $\Phi$-values for mutations in the helix of protein L. 
The solid line represents the regression line $\Phi = 0.06 + 0.24 \,\Delta G_\alpha/\Delta G_N$ for all data points; the dotted line is the regression line $\Phi = 0.15 + 0.06 \,\Delta G_\alpha/\Delta G_N$ of the 7 data points for mutations that affect tertiary interactions with hairpin 1 or none of the tertiary interactions (see also Table 4); and the dashed line is the regression line $\Phi = -0.04 + 0.37 \,\Delta G_\alpha/\Delta G_N$ of the 6 data points for mutations affecting tertiary interactions with hairpin 2 or none of the tertiary interactions. } 
\label{helix_proteinL}
\end{figure}

The structural parameters $\chi_\alpha$ and $\chi_t$ obtained from our analysis shown in the Figs.~\ref{helix_ci2} to \ref{helix_proteinL} are summarized in Table 5. We estimate the overall errors of  $\chi_\alpha$ and $\chi_t$, which result from experimental errors in $\Phi$ and $\Delta G_N$ and from modeling errors, as $\pm 0.05$ for the CI2 helix and helix 2 of protein A, and as $\pm 0.1$ for helix 3 of protein A and the protein L helix. The $\chi_\alpha$ values for the CI2 helix and the helix 2 of protein A are close to 1. This indicates that the helices are fully formed in the transition-state ensemble. In contrast, $\chi_\alpha$ for helix 3 of protein A is close to 0, indicating that the helix is not formed in the transition state. $\chi_\alpha$ for the helix in protein L indicates a partial degree of helix formation between 20 and 30 \%. Our  $\chi_t$ values indicate that the degree of tertiary structure formation in the transition state is around 16 \% for the CI2 helix, around 50 \% for helix 2 of protein A, and around 30 \% for helix 3 of protein A. The $\chi_t$ values for the protein L helix show a small degree of tertiary structure formation with hairpin 1 (around 15 \%) and no tertiary structure formation with hairpin 2. 

To assess the quality of our modeling, we consider two quantities: the correlation coefficient $r$, and the estimated standard deviation SD of the data points from the regression line. High correlation coefficients up to 0.9 and larger indicate a high quality of modeling. However, it's important to note that the correlation coefficient can only be used to assess the modeling quality in the cases where the structural parameters $\chi_\alpha$ and $\chi_t$ are sufficiently different from each other. The case $\chi_\alpha = \chi_t$ corresponds to a regression line with slope 0, and hence a correlation coefficient of 0, irrespective of how well the data are represented by this line. For small differences of $\chi_\alpha$ and $\chi_t$, the correlation coefficient $r$ is dominated by the experimental errors in $\Phi$.  This is the case for the mutations in the protein L helix that affect tertiary contacts with hairpin 1, see Table 5.  The slope of the regression line is almost zero for this data set, see Fig.~\ref{helix_proteinL}. Here, the relatively small standard deviation 0.1 of the data points from the regression lines indicates that our model is in good agreement with the experimental data. 

We only consider here mutations with stability changes $\Delta G_N>0.7$ kcal/mol. Because of experimental errors, $\Phi$-values for mutations with smaller stability changes are generally considered as unreliable \cite{Fersht04,Garcia04,Rios06}. In our previous work \cite{Merlo05}, we considered all the published mutations for the CI2 helix, including those for which $\Delta G_N$ is significantly smaller than 0.7 kcal/mol. The correlation coefficient 0.91 obtained here for the subset of mutations with $\Delta G_N>0.7$ kcal/mol is larger than the correlation coefficient 0.85 for all mutations. The significantly larger reliability threshold of 1.7 kcal/mol for $\Delta G_N$ obtained by Sanchez and Kiefhaber \cite{Sanchez03} is based on the assumption that different mutations at the same residue position should lead to the same $\Phi$-value. In our model, different $\Phi$-values for mutations at the same site result from different effects on the intrinsic helix stability $G_\alpha$ and the tertiary free energy $G_t = G_N - G_\alpha$.

In our model, nonclassical $\Phi$-values $<0$ or $>1$ can arise if $\Delta G_\alpha/\Delta G_N$ is $<0$ or $>1$. Since $\Delta G_N = \Delta G_\alpha +  \Delta G_t$, this implies that $\Delta G_\alpha$ and $\Delta G_t$ have opposite signs. Our model reproduces the clearly negative $\Phi$-values for the mutations D23A in the CI2 helix and D38A in the protein L helix. Both mutations stabilize the helix (i.e.~$\Delta G_\alpha<0$), but  destabilize tertiary interactions ($\Delta G_t>0$).
 \footnote{In our previous article \cite{Merlo05}, we had erroneously stated that nonclassical $\Phi$-values can arise for mutations that only shift the free energy of the denatured state, but not the free energy of the transition state and native state. This is not the case. Indeed, the $\Phi$-value for these hypothetical mutations is 1 since $\Delta G_T=\Delta G_N$.}

\section*{Conclusions}
We have shown how to obtain a structural interpretation of $\Phi$-values for multiple mutations within a single helix.  Combined with any scale of helical propensities, our model shows how linear fitting of experimental data leads to two structural quantities: the extent of helix formation in the transition state, and the extent to which the helix interactions with neighboring tertiary structure are formed in the transition state.  The method gives a simple physical interpretation of nonclassical $\Phi$-values -- nonclassical values arise if a mutation stabilizes a helix while destabilizing its interactions with neighboring parts of the protein, or vice versa.  The model also explains how two different mutations at the same site can have different effects on the kinetics -- this difference is traced back to different effects of the mutations on the intrinsic helix stability versus tertiary stability. Hence, this model appears to give simple physically consistent structural explanations for experimentally measured $\Phi$-values.

\clearpage

\begin{table}
\hspace*{2cm}
Table 1: Helix of the protein CI2\\ 
\begin{center}
\begin{tabular}{ccccc}
mutation & $\Phi$ & $\Delta G_N$   & $\Delta G_\alpha^\text{AGADIR} $ & $ \Delta G_\alpha^\text{prop}$ \\ 
\hline
S12G & 0.29 & 0.8 & 0.28 &  -- \\
S12A & 0.43 & 0.89 & 0.14 & -- \\
E15D & 0.22 & 0.74 & 0.13 & 0.29 \\
E15N & 0.53 & 1.07 & 0.57 & 0.25 \\
A16G & 1.06 & 1.09 & 0.82 & 1.0 \\
K17G & 0.38 & 2.32 & 0.80 &  0.74 \\
K18G & 0.7 &   0.99 & 0.75 &  0.74 \\
I20V & 0.4 &     1.3 &   0.14 & 0.2 \\
L21A & 0.25 & 1.33 &  -0.01 &  -0.21 \\
L21G & 0.35 & 1.38 &  0.26 & 0.79 \\
D23A & -0.25 & 0.96 &  -0.41 & --  \\
K24G & 0.1 &   3.19 &  0.12 &  -- 
\end{tabular}
\vspace{0.5cm}
\end{center}
Experimental $\Phi$-values and stability changes $\Delta G_N$ are from Itzhaki et al.\cite{Itzhaki95}. The change in intrinsic helix stability $\Delta G_\alpha^\text{AGADIR}$ is calculated with AGADIR \cite{Munoz94a,Munoz94b,Lacroix98}, see Merlo et al.~\cite{Merlo05}. The change in intrinsic helix stability $\Delta G_\alpha^\text{prop}$ is calculated from the helix propensity scale of Pace and Scholtz \cite{Pace98}. The helix propensities of the residues are (in kcal/mol): Ala (A) 0, Leu (L) 0.21, Arg (R) 0.21, Met (M) 0.24, Lys (K) 0.26, Gln (Q) 0.39, Glu (E) 0.40, Ile (I) 0.41, Trp (W) 0.49, Ser (S) 0.50, Tyr (Y) 0.53, Phe (F) 0.54, Val (V) 0.61, His (H) 0.61, Asn (N) 0.65, Thr (T) 0.66, Cys (C) 0.68, Asp (D) 0.69, and Gly (G) 1. For the terminal residues 12, 13, 23, and 24 of the helix, the propensity scale is not applicable. We only consider mutations with $\Delta G_N> 0.7$ kcal/mol. 
\end{table}


\begin{table}
\hspace*{2cm}
Table 2: Helix 2 of protein A\\ 
\begin{center}
\begin{tabular}{ccccc}
mutation & $\Phi$ & $\Delta G_N$ & $\Delta G_\alpha $ & tertiary contacts\\ 
 \hline
A27G & 1.0 & 1.0 & 1.0 & -- \\ 
A28G & 0.6 & 2.2 & 1.0 & Helix 1 \\ 
A29G & 1.1 & 1.0 & 1.0 & -- \\
F31A & 0.3 & 3.9 & -0.54 & Helices 1, 3 \\ 
F31G & 0.5 & 4.7 & 0.46 & Helices 1, 3 \\ 
I32V & 0.6 & 1.2 & 0.2 & Helix 1\\ 
I32A & 0.5 & 1.9 & -0.41 & Helix 1\\ 
I32G & 0.6 & 3.4 & 0.59 & Helix 1\\ 
A33G & 1.1 & 0.9 & 1.0 & -- \\ 
A34G & 0.7 & 1.2 & 1.0 &  -- \\ 
L35A & 0.4 & 2.4 & -0.21 & Helices 1, 3  \\ 
L35G & 0.5 & 4.1 & 0.79 & Helices 1, 3 
\end{tabular}
\vspace*{0.5cm}
\end{center}
Experimental $\Phi$-values and stability changes $\Delta G_N$ are from Sato et al.\ \cite{Sato04}. The change in intrinsic helix stability $\Delta G_\alpha$ is calculated from the helix propensity scale of Pace and Scholtz \cite{Pace98}. The information whether tertiary contacts with helix 1 and 3 are affected by the mutations is taken from the contact matrix of protein A shown in Fig.~\ref{map_proteinA}. We only consider $\Phi$-values for single-residue mutations with the wildtype sequence as reference state at those sites where multiple mutations have been performed. For example, we consider the $\Phi$-values for the mutations I32V, I32A, and I32G in helix 2 of protein A, but not the $\Phi$-values for V32A and A32G also given by Sato et al.\ \cite{Sato04}. However, we include the $\Phi$-values for the Ala-Gly scanning mutants at the residue positions 27, 28, 29, 33, and 34 given in Table 1 of Sato et al.\ \cite{Sato04}.
\end{table}

\clearpage

\begin{table}
\hspace*{2cm}
Table 3: Helix 3 of protein A\\ 
\begin{center}
\begin{tabular}{ccccc}
mutation & $\Phi$ & $\Delta G_N$ & $\Delta G_\alpha $ & tertiary contacts\\ 
 \hline
A44G & -0.1 & 1.3 & 1.0 &  -- \\
L45A & 0.6 & 1.5 & -0.21  & Helix 2 \\
L45G & 0.3 & 4.4 & 0.79  & Helix 2 \\
L46A & 0.2 & 1.9 & -0.21  & Helix 1 \\
L46G & 0.3 & 4.0 & 0.79  & Helix 1 \\
A47G &  0.2 &  1.5 &  1.0  & -- \\
A48G &  0.0 &  1.8 &  1.0 & Helix 2 \\
A49G &  0.2 &  3.6 &  1.0  & Helix 2\\
A51G &  0.1 &  1.2 &  1.0 & -- \\
L52A &  0.3 &  1.3 &  -0.21 & Helix 2 \\
L52G &  0.1 &  3.8 &  0.79 & Helix 2 \\
A54G &  0.0 &  1.4 &  1.0 & -- 
 \end{tabular}
 \vspace{0.5cm}
\end{center}
Experimental $\Phi$-values and stability changes $\Delta G_N$ are from Sato et al.\ \cite{Sato04}. The change in intrinsic helix stability $\Delta G_\alpha$ is calculated from helix propensities \cite{Pace98}. The information on tertiary contacts is taken from Fig.~\ref{map_proteinA}
\end{table}

\clearpage

\begin{table}

\hspace*{2cm}Table 4:  Helix of protein L\\ 

\begin{center}
\begin{tabular}{ccccc}
mutation & $\Phi$ & $\Delta G_N$   & $\Delta G_\alpha$  & tertiary contacts\\ 
 \hline
A29G & 0.23 & 2.41 & 1.0 &  Hairpin 1 \\
T30A & 0.08 &  1.31 & -0.66 & Hairpin 1 \\
S31G & 0.11 & 0.81 & 0.5  & -- \\
E32G & 0.11 & 1.08 & 0.6 & Hairpin 1 \\
E32I & 0.05 & 1.25 & 0.01 & Hairpin 1 \\
A33G & 0.25 & 2.85 & 1.0 & Hairpin 1, 2 \\
Y34A & 0.05 & 2.57 & -0.53 & Hairpin 2 \\
A35G & 0.28 & 1.2 & 1.0 & -- \\
Y36A & 0.27 & 2.54 & -0.53 & Hairpin 1 \\
A37G & 0.11 & 3.14 & 1.0 & Hairpin 2 \\
D38A & -0.39 & 0.98 & -0.69 & Hairpin 2 \\
D38G & -0.05 & 1.89 & 0.31 & Hairpin 2 
\end{tabular}
\vspace*{0.5cm}
\end{center}
Experimental $\Phi$-values and stability changes $\Delta G_N$ are from Kim et al.\ \cite{Kim00}. The change in intrinsic helix stability $\Delta G_\alpha$ is calculated from helix propensities \cite{Pace98}. The two $\beta$-hairpins of protein L are defined in the caption of Fig.~\ref{map_proteinL}. The information on tertiary interactions of helical residues with the hairpins is taken from this figure.
\end{table}

\clearpage

\begin{table}

Table 5: Structural parameters, standard deviations, and correlation coefficients \\

\begin{center}
\begin{tabular}{c|c|cc|cc}
helix  & tertiary contacts & $\chi_\alpha$ & $\chi_t$ & SD &$|r|$ \\ 
 \hline
CI2 helix & all & 1.03 &  0.16 &  0.14 & 0.91 \\    
\hline   
helix 2 of protein A & all & 0.98 & 0.46 &  0.10  & 0.93 \\
                & with helix 1 & 0.98 & 0.52 & 0.12  & 0.90 \\
\hline
helix 3 of protein A & all & -0.07 & 0.31 & 0.13  & 0.75 \\
                 & with helix 1 &  -0.01 & 0.24 &  0.13 & 0.65 \\
                 & with helix 2 &  -0.09 & 0.34 & 0.13  & 0.79 \\
\hline
helix of protein L & all  &  0.30 & 0.06 & 0.15  & 0.63 \\
                &  with hairpin 1 & 0.21 & 0.15 & 0.10 & (0.30)$^a$ \\
             & with hairpin 2 & 0.32 & -0.04 &  0.11 & 0.90 \\
\end{tabular}
\vspace*{0.5cm}
\end{center}
The structural parameters $\chi_\alpha$ and $\chi_t$, estimated standard deviations SD of the data points from the regression lines, and absolute values of the correlation coefficient $r$ obtained in our model. The second column of the table indicates whether we consider all mutations for a helix, or only mutations affecting tertiary interactions with one structural element. The structural parameter $\chi_t$ then either indicates the overall degree of tertiary structure formation in the transition state, or the degree of tertiary structure formation with the given structural element. In both cases, we have included the `purely secondary' mutations that do not affect tertiary interactions. The structural elements of protein A and L are defined in the Figs.~\ref{map_proteinA} and \ref{map_proteinL}.  The standard deviation SD is estimated as SD = $\sqrt{\left(\sum_{i=1}^M d_i^2\right)/(M-2)}$ where $d_i$ is the vertical deviation of data point $i$ from the regression line, and $M$ is the number of data points. We estimate the errors in the structural parameters $\chi_\alpha$ and $\chi_t$, which result from experimental and modeling errors, as $\pm 0.05$ for the CI2 helix and helix 2 of protein A, and as $\pm 0.1$ for helix 3 of protein A and the protein L helix.
\\
$^a$For this data set, the correlation coefficient $r$ is not a reasonable indicator of the modeling quality since the slope of the regression line is close to 0. The precise value of $r$ is then dominated by the experimental errors in $\Phi$. In our model, the slope of the regression line close to 0 indicates that the two structural parameters $\chi_\alpha$ and $\chi_t$ have similar values, see eq.~\ref{phi_modeled}. The relatively small standard deviation SD of 0.10 for this data set shows that our model is in good agreement with the data.
\end{table}

\clearpage

\clearpage

\clearpage

\clearpage

\clearpage

\clearpage

\clearpage

\end{document}